\documentclass[prb,twocolumn,showpacs]{revtex4}%
\usepackage{amsmath}
\usepackage{amsfonts}
\usepackage{amssymb}
\usepackage{graphicx}%
\begin{document}
\title{Constant effective mass across the phase diagram of high-T$_{c}$ cuprates}
\author{W. J. Padilla}\altaffiliation{Present address: Los Alamos National
Laboratory, MS G756, MST-CINT, Los Alamos, NM 87545.}
\email{willie@lanl.gov}
\author{Y. S. Lee}
\author{M. Dumm}\altaffiliation{Present address: 1. Physikalisches Institut, Universit\"{a}t Stuttgart, 70550
Stuttgart, Germany} \affiliation{Department of Physics, University
of California at San Diego, La Jolla, CA 92093-0319.}
\author{G. Blumberg}
\affiliation{Bell Laboratories, Lucent Technologies, Murray Hill, New Jersey 07974, USA.}
\author{S. Ono}
\author{Kouji Segawa}
\author{Seiki Komiya}
\author{Yoichi Ando}
\affiliation{Central Research Institute of Electric Power Industry, Komae, Tokyo 201-8511, Japan.}
\author{D. N. Basov}
\affiliation{Department of Physics, University of California at San Diego, La Jolla, CA 92093-0319}


\begin{abstract}
We investigate the hole dynamics in two prototypical high
temperature superconducting systems: La$_{2-x}$Sr$_{x}$CuO$_{4}$ and YBa$_{2}$Cu$_{3}%
$O$_{y}$ using a combination of DC transport and infrared
spectroscopy. By exploring the effective spectral weight obtained
with optics in conjunction with DC Hall results we find that the
transition to the Mott insulating state in these systems is of the
``vanishing carrier number" type since we observe no substantial
enhancement of the mass as one proceeds to undoped phases. Further,
the effective mass remains constant across the entire underdoped
regime of the phase diagram. We discuss the implications of these
results for the understanding of both transport phenomena and
pairing mechanism in high-T$_{c}$ systems.

\end{abstract}

\maketitle

At zero temperature in a Mott-Hubbard (MH) insulator, carriers are
localized due to strong electron-electron interactions. In some
systems long range antiferromagnetic (AF) order is favored due to
superexchange.\cite{anderson} Moving away from this ground state
by heating or doping, a number of exotic behavior
emerge.\cite{imada,kotlier} These include unconventional magnetic
order\cite{yamada}, pseudogap phenomena\cite{timusk}, and
high-T$_{c}$ superconductivity (SC). To elucidate these complex
forms of matter, it is necessary to clarify the origins of
conduction as evolved from the undoped zero temperature state of
the MH insulator. In fact, since the discovery of high-T$_{c}$ SC
this task has been widely regarded as one of the most fundamental
problems in all of condensed matter physics.\cite{orenstein} Here
we study the only known class of MH systems to exhibit
high-T$_{c}$ superconductivity, the cuprates. Transport and
infrared properties are presented for two prototypical families:
La$_{2-x}$Sr$_{x}$CuO$_{4}$ (LSCO) and YBa$_{2}$Cu$_{3}$O$_{y}$
(YBCO). We show that the optical effective mass remains constant
throughout the underdoped region of the phase diagram and discuss
the implications of this enigma for both transport and
superconductivity.

As a probe into the carrier dynamics of these two archetypal
compounds we utilize a combination of DC transport and infrared
spectroscopy. The optical measurements have been carried out from
1.2 meV to 6 eV (10 cm$^{-1}$ to 50,000 cm$^{-1}$) in near normal
reflectance geometry. Both the LSCO and YBCO samples are high
quality de-twinned single crystals, the preparation of which was
described previously in detail.\cite{ando} Detwinned samples are
especially important for YBCO to ensure that the a-axis data
analyzed in this work are not contaminated by the contribution due
to Cu-O chain segments. The optical conductivity
$\sigma_{1}(\omega)+i\sigma_{2}(\omega)$\ and complex dielectric
function $\varepsilon_{1}(\omega)+i\varepsilon_{2}(\omega)$\ was
determined from the reflectance data\cite{wooten,woot} and checked
against direct ellipsometric measurements. This work encompasses
the underdoped regime of the phase diagram from long range AF
ordered insulators, to high-T$_{c}$ SC
compounds.%

\begin{figure*}
[ptb]
\begin{center}
\includegraphics[
height=4.024in,
width=6.7006in
]%
{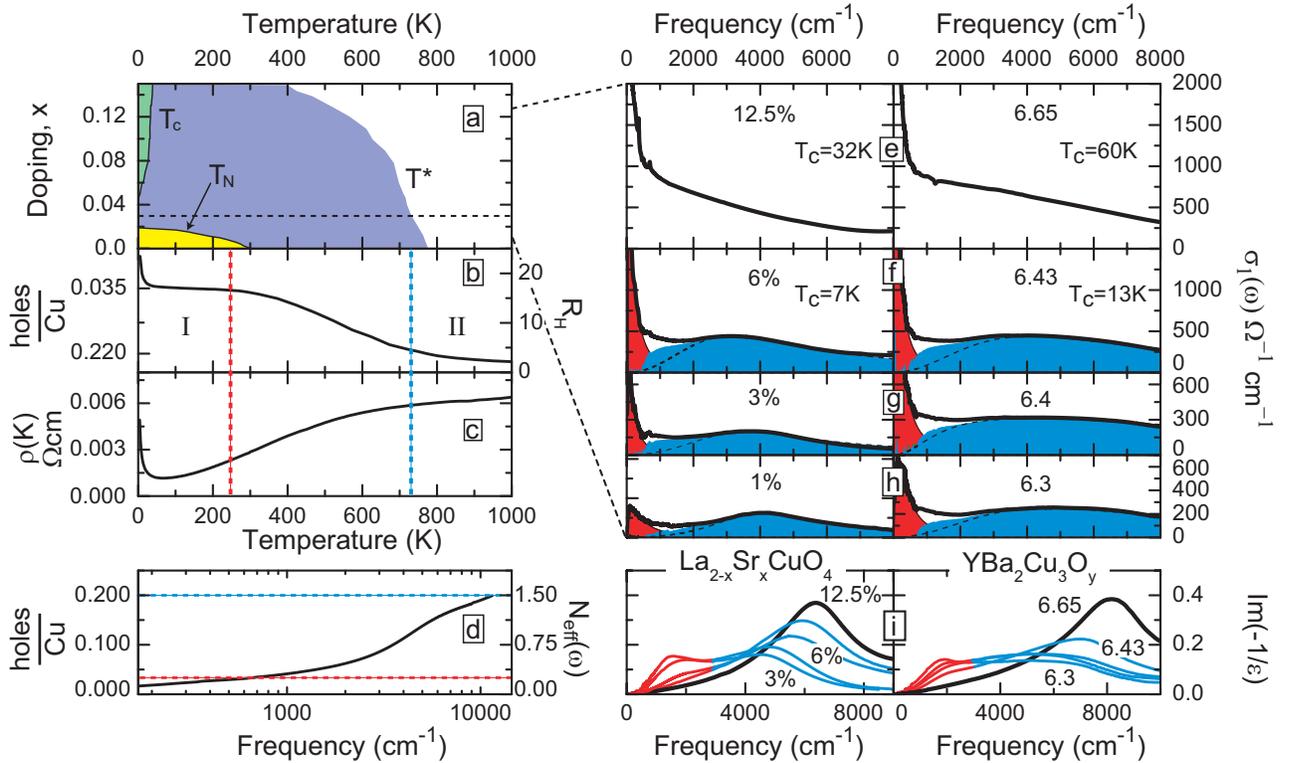}%
\caption{(Color online). Optical data for YBCO and both optical
and transport data for LSCO discussed in the context of the
high-T$_{c}$ phase diagram. Left panels are data for x=0.03 LSCO
and right panels depict the low temperature optical conductivity
as a function of doping for both LSCO and YBCO. Data in panels f-h
are at T= 7 K and at T just above T$_c$ in panels e. Panel a:
phase diagram with the values shown for LSCO; the yellow area is
the long range AF region (denoted by T$_{N}$),the green area:
superconducting region (marked by T$_{c}$), and the pseudogap
region is denoted by T*. To facilitate a direct comparison with
the data, temperature is plotted on the horizontal axis and doping
on the vertical axis. The Hall coefficient R$_{H}$ (10$^{-3}$
cm$^{3}$/C) displayed in panel (b) reveals two characteristic
regimes (I and II). This is exemplified for the x=0.03 sample but
similar behavior is found for all heavily underdoped crystals.
These regimes can also be identified with significant regions in
the DC resistivity corrected for thermal expansion (c). Panel d:
the effective number of carriers, N$_{eff}$ (10$^{6}$
$\Omega^{-1}$cm$^{-2}$), determined from the optical data as
described in the text. Right panels: the real part of the
conductivity at low temperature (T=10K, except T=T$_{c}$ for e and
f and T=150K for y=6.65), displayed as black solid curves. Red
colors signify fits using the standard Drude model. The blue area
is a subtraction of the Drude fit from $\sigma_{1}$($\omega$) and
is referred to as the \textquotedblleft mid-infrared
contribution\textquotedblright\ throughout the text. The resonant
structure in the latter contribution can be adequately described
with a Lorentzian oscillator (dashed lines). The loss function
[Im(-1/$\varepsilon$)], panels (i), also clearly depicts two
components (shown as red and blue curves) for dopings x$<$0.10,
y$<$6.65, and the progression to a single component, (black solid
curve) for higher dopings.}%
\label{fig1}%
\end{center}
\end{figure*}

\begin{figure}
[ptb]
\begin{center}
\includegraphics[
height=2.3194in, width=3.2897in
]%
{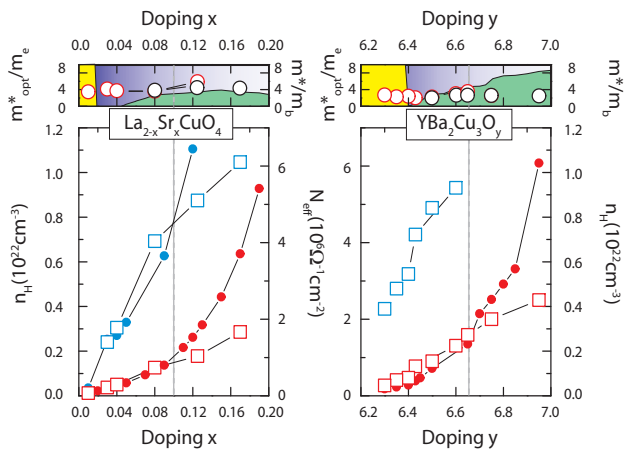}%
\caption{(Color online). Top panels show the optical effective
mass as determined by the methods described in the text. Red
symbols are the mass determined using a combination of optics and
transport (left axis labels), and the black symbols represent the
mass determined within the extended Drude model (right labels).
All methods produce consistent results and mass values which are
doping independent over large regions of the phase diagram. The
low temperature long range AF ordered phase is denoted by the
yellow shaded area and SC region by the green area. Note that
despite the formation of the AF regime near the Mott transition,
there is no noticeable enhancement of the optical mass. Bottom
panels show both the effective spectral weight from optics data
(red and blue open squares) and the number of holes as determined
by transport
(red and blue dots).}%
\label{fig2}%
\end{center}
\end{figure}

The real part of the conductivity $\sigma_{1}(\omega)$\ is
displayed in Fig. \ref{fig1} (panels e-h) and shows the evolution
of the doping trends. It has been established that the lowest
electronic excitation in undoped crystals of different families of
high-T$_{c}$ superconductors is associated with a charge transfer
(CT) gap at $\simeq$12000 cm$^{-1}$=1.5eV.\cite{tanner} The data
presented in Fig. \ref{fig1} reveals the development of two
distinct features in the intra-gap response upon carrier doping,
in accord with the earlier results.\cite{tanner,uchida} The
low-energy response is dominated by a Drude-like coherent
contribution to the conductivity followed by a broad resonance in
the mid-IR. We represent the dissipative part of the conductivity
composed of two components as, $\sigma_{1}(\omega)=\omega_{pD}^{2}\frac{\gamma_{D}}%
{\omega^{2}+\gamma_{D}^{2}}+\omega_{pM}^{2}\frac{\gamma_{M}\omega^{2}}%
{(\omega_{M}^{2}-\omega^{2})+\gamma_{M}^{2}\omega^{2}}$, where $\omega_{p}%
^{2}$ is the square of the plasma frequency for each term
(subscripts D and M are for the Drude and MIR terms respectively),
$\omega_{M}$ is the center frequency of the MIR term, $\gamma_{D}$
and $\gamma_{M}$ are the Drude and MIR damping terms respectively.
With increasing doping the oscillator strength associated with
both contributions is enhanced (this is further detailed in Fig.
\ref{fig2}). The low-energy contribution to the conductivity is
adequately described with the Drude expression (red regions in
Fig. \ref{fig1} f-h) revealing the metallic nature of the
electronic transport even in AF-ordered
phases.\cite{footnoteloc,ando2,ando} Approaching optimal doping at
$T=T_c$, the two components in the conductivity merge and can no
longer be unambiguously identified. Notably, both the LSCO and
YBCO series reveal exactly the same trends.

New insights into the nature of electronic transport in underdoped
phases can be inferred from a combination of the $\sigma(\omega)$
results discussed above and the Hall and resistivity data obtained
for nominally identical single crystals. We first note that in
very underdoped crystals the behavior of the Hall coefficient is
rather simple.\cite{ando2} This is exemplified for x=0.03 crystal
(Fig. \ref{fig1}b). Indeed, $R_H^{-1}$ is essentially independent
of temperature below 300-400 K until the lowest temperatures where
both transport and optics data are strongly influenced by Anderson
localization.\cite{scale} Moreover, the carrier density $n_H=
R_H^{-1}$ extracted from this plateau is in fair agreement with an
estimate based on the chemical composition of 0.035 holes per
copper atom.\cite{ando,ando2} The latter inference holds for other
underdoped phases of both the LCCO and YBCO
series.\cite{ando,segawa}

In order to establish parallels between transport and optics
results it is customary to define the effective number of carriers
$N_{eff}(\omega)=\int_{0}^{\omega}d\omega^{\prime}\sigma_{1}(\omega^{\prime})$
contributing to the conductivity below a cut-off frequency
$\omega$. The strength of the coherent contribution to the
conductivity can be quantified either through the parameters of
the Drude fit or by limiting the integration cut-off to 650
cm$^{-1}$. The results of the latter approach are displayed with
the filled red symbols in Fig. \ref{fig2}. The coherent
contribution amounts to approximately 20$\%$ of the total spectral
weight below the CT excitation (Fig. \ref{fig1}d and open blue
squares in Fig. \ref{fig2}). Inspecting the data in Fig.
\ref{fig2} one finds that both $n_H$ and $N_{eff}^{coh}$ reveal
the same doping dependence in underdoped compounds where two
component response can be identified in the IR data. With doping
progressing towards optimal one witnesses a departure between the
two results, again occurring in the same fashion for both the LSCO
and YBCO series.

Encouraged by the close consistency between optics and Hall data
in underdoped regime we evaluated the optical effective mass of
mobile holes as $m_{Opt.}^{\ast}=(R_{H}\times
N_{eff}^{coh})^{-1}$. These mass values are plotted as the red
symbols in the top panel of Fig. \ref{fig2}. At doping levels
above a critical value (x$\sim$0.10 for LSCO and y$\sim$6.65 for
YBCO) one cannot unambiguously separate the coherent component
from the mid-IR background. Nevertheless, the mass enhancement
over the band value can still be determined using solely optical
data within the frame-work of the \textquotedblleft extended Drude
model\textquotedblright\: $\frac{m^{\ast}}{m_{b}}=-\frac{\omega_{p}^2}%
{4\pi}\frac{1}{\omega}Im[\frac{1}{\sigma(\omega)}]$ (black symbols
in top panel of Fig. 2),\cite{puchkov} and these are plotted as
the black symbols in Fig. \ref{fig2}. Interestingly the two
approaches produce consistent results. The key outcome of this
analysis is that $m_{Opt.}^{\ast}$ is nearly doping independent in
both LSCO and YBCO.

The topology of the Fermi surface in weakly doped cuprates is such
that the masses calculated here reveal important dynamical
characteristics of the electronic states in the nodal region, and
are thus related to the nodal quasiparticle effective mass. Doping
independent $m_{Opt.}^{\ast}$ in both families of cuprates
excludes the notion that carriers doped in MH system necessarily
experience a divergence of $m^{\ast }$ near the transition to an
insulator.\cite{brinkman} Although we are not aware of any direct
inference of the effective mass for the cuprates from the specific
heat data, an inspection of the doping dependence of linear
coefficient $\gamma$ supports our findings for
$m_{Opt}^*$.\cite{kumagai,loram}

A theoretical analyses of the dynamics of mobile holes introduced
into an antiferromagnet with the exchange energy \textit{J}
suggests that the mass divergence can be avoided.\cite{trugman,
patrick, sachdev} Nevertheless, the effective mass is still
strongly renormalized by the ratio of \textit{t}/\textit{J} where
\textit{t} is the bandwidth. A common denominator in these models
is that the motion of a hole in an AF background implies a
continuous reorganization of the local spin environment. Since
these processes are energetically costly they restrict carrier
transport and enhance the effective mass with\textit{
t}/\textit{J}$\sim$10 in the Cuprates.\cite{patrick} The
experimental situation for non-superconducting MH oxides is
somewhat controversial. For example some materials indeed comply
with the above expectations and show a strong increase of
$m^{\ast}$ when the Mott insulator boundary is approached both by
varying concentration of dopants\cite{imada,tokura,tokura2} and
hydrostatic pressure.\cite{carter} Other systems do not show such
an enhancement.\cite{katsufuji} Our data for high-T$_{c}$ cuprates
report, for the first time, a systematic study of the optical
effective mass in this class of materials, and reveal relatively
light masses. More importantly no noticeable changes in
$m_{Opt.}^{\ast}$ are found despite the formation of long range AF
ordered phases, and even in immediate proximity to the Mott
insulating state.

It is instructive to discuss the constant effective mass result in
conjunction with other universal trends of cuprates. Our
observations imply that the development of the conducting state
occurs primarily through an increase in the density of carriers,
while the dynamical characteristics of mobile holes remain
remarkably constant throughout the entire phase diagram. This
conclusion is in accord with transport data\cite{kumagai,ando} and
also supported by recent photoemission studies.\cite{shen} The
latter work reports a gradual development of the Fermi arc and
finds that the integrated weight varies proportionally to $n$ with
doping. Interestingly, measurements along the ``nodal region" (the
direction of [0,0] to [$\pi,\pi$] in k-space), find that the Fermi
velocity also is insensitive to doping.\cite{zhou} One conjecture
reconciling all these observations is that the local environment
of mobile charges in cuprates remains unaltered with doping and it
is only the phase space occupied by hole rich regions that is
progressively increasing. Some of these trends are also consistent
with the variational analysis of doped holes in a resonant valence
bonds insulator.\cite{paramekanti}

We now wish to highlight several unexpected features in the high
temperature DC transport of very underdoped crystal relating these
features to the IR results. The Hall number $n_{H}=R_H^{-1}$ shows
a marked increase at $T>300-400$ K with a new plateau emerging
above 800-900 K as exemplified for x=0.03 LSCO crystal in Fig.
\ref{fig1}b. An increase in the Hall number occurs primarily
between regions I and II (Fig. \ref{fig1} b) and appears to impact
the T dependence of the resistivity $\rho(T)$, Fig. \ref{fig1}\
(c). Indeed the rapid growth of the resistivity with T in regime I
is followed by a nearly T-independent resistivity in regime
II.\cite{resist} The inverse resistivity in the saturated region
is in quantitative agreement with the magnitude of conductivity in
at mid-IR frequencies. It is known that the coherent Drude-like
feature clearly seen in the low-T data in Fig. \ref{fig1}
significantly broadens with increasing T and can no longer be
discriminated from the incoherent mid-IR background as T is
elevated to ~400 K.\cite{takenaka} We speculate that the
transformation of $\sigma(\omega)$, resistivity and $n_H$ at
elevated temperatures may all be of common origin. At these high
temperatures quasiparticles residing on the Fermi arcs loose their
identity and it is no longer possible to distinguish the role of
these excitations in the transport properties from that of an
``incoherent background". One corollary of the results reported
here is that the magnitude of $m_{Opt.}^{\ast}$ reported in Fig.
\ref{fig2} remains unchanged if this value is extracted from the
total spectral weight (below the charge transfer absorption) in
combination with $n_H$ at 800-900 K.

\qquad In conclusion, we have investigated the optical mass
associated with the response of nodal quasiparticles across the
underdoped regime of two prototypal high-T$_{c}$ superconducting
families. Our results establish an alternative route for the Mott
transition in this class of materials without the usual mass
divergence or enhancement. One implication of a constant effective
mass is that transport in high-T$_{c}$ cuprates is governed by
excitations topologically compatible with an antiferromagnetic
background. Specific schemes permitting a constant effective mass
include spin-charge separation\cite{patrick} and electronic phase
separation.\cite{emery3} Both scenarios propose unconventional and
exotic explanations of high-T$_{c}$ superconductivity which depart
from the standard BCS prescription.


\begin{thebibliography}{99}
\bibitem {wooten}This was obtained by a Kramer-Kronig transformation of the
data after extrapolation to low and high energies.
\bibitem {footnoteloc}  For x=0.01 and 0.03 LSCO crystals we used
localization modified Drude formula accounting for a shift of the
Drude feature to finite frequency: a hallmark of Anderson
localization in $\sigma(\omega)$. While Anderson localization
modifies the response of the system at lowest frequencies it is
nevertheless appropriate to classify these crystals as "metallic".
Notably, a significant electronic spectral weight extends down to
the lowest energies without any indications for a gap in the
density of states. The AF-ordered YBCO crystal with y=6.3 reveals
a conventional Drude response, at lower dopings we observed a
Anderson localization feature in YBCO compounds as well.
\bibitem {scale}Both the resistivity and the Hall data for very light dopings
are influenced by carrier localization [upturns in both R$_{H}$(T)
and $\rho $(T)]. Localization trends are also evident in the
$\sigma$($\omega$).  This behavior is not analyzed further here.
\bibitem {resist}Although the resistivity shown in Fig. \ref{fig1}(c)
continues to slowly increase between regions I and II, the
curvature inflection point occurs at approximately 300K. Thus the
onset of resistivity saturation occurs simultaneously with the
increase in the number of carriers.
\end{thebibliography}
\end{document}